\documentclass[a4paper,10pt,twoside]{cpc-hepnp}

\usepackage{multicol}
\usepackage{graphicx}
\usepackage{booktabs}
\usepackage{amssymb,bm,mathrsfs,bbm,amscd}
\usepackage[tbtags]{amsmath}
\usepackage{lastpage}
\usepackage{float}

\begin{document}

\fancyhead[c]{\small Chinese Physics C~~~Vol. 37, No. 1 (2013)
010201} \fancyfoot[C]{\small 010201-\thepage}

\footnotetext[0]{Received 14 March 2012}

\title{Impact parameter dependence of the scaling of anisotropic flows in intermediate energy HIC
}

\author{YAN Ting-Zhi(ÑÕÍ¢Ö¾)$^{1;1)}$\email{ytz0110@163.com}
\quad LI Shan(Àîɼ)$^{2}$
}

\maketitle

\address{%
$^{1}$ School of Energy Resources and Power Engineering, Northeast
Dianli University, Jilin 132012, China\\
$^2$ School of Science, Northeast
Dianli University, Jilin 132012, China\\
}

\begin{abstract}
The scaling behaviors of anisotropic flows of light charged
particles are studied for 25 \,MeV/nucleon $^{40}$Ca+$^{40}$Ca
collisions at different impact parameters by the isospin-dependent
quantum molecular dynamics model. The number of nucleons scaling
of elliptic flow is existed and the scaling of the ratios of
$v_{4}/v_{2}^{2}$ and $v_{3}/(v_{1}v_{2})$ are applicable for
collisions at almost all impact parameters except for peripheral
collisions.
\end{abstract}

\begin{keyword}
Impact parameter, anisotropic flows, scaling behaviors
\end{keyword}

\begin{pacs}
25.75.Ld, 24.10.-i, 21.60.Ka
\end{pacs}

\begin{multicols}{2}

\section{Introduction}

Anisotropic flows are interesting subjects in theoretical and
experimental investigations on nuclear reaction dynamics in both
intermediate and high energy heavy ion collisions. Many studies of
the first and second anisotropic flows (directed flow and elliptic
flow, respectively) dependence on beam energies, mass number or
quark number, isospin and impact parameter have been done and
revealed much interesting physics about the properties and origin
of the collective flow\cite{J. Ollitrault,H. Sorge,P.
Danielewicz,D. Teaney,P. F. Kolb,Y. Zheng,D. Perslam,J.
Lukasik,J.Adams,Ma06}. $^{197}$Au + $^{197}$Au collision
experiments at RHIC energy demonstrated the number of
constituent-quark (NCQ) scaling for the transverse momentum
dependent elliptic flow for different mesons and
baryons\cite{Ko,Molnari}, and a popular interpretation is assuming
that the mesons and baryons are formed by the coalescence or
recombination of the constituent quarks. Our work\cite{Yan} found
the similar elliptic scaling for light particles in intermediate
energy heavy ion collisions and it may also be the outcome of
coalescence mechanism but at nucleonic level. As we know, the flow
value is strongly depended on the impact parameter, so the
elliptic scaling is tested for light particles at different impact
parameter of $^{40}$Ca + $^{40}$Ca collisions in intermediate
energy in this paper, and $v_{4}/v_{2}^{2}$ and
$v_{3}/(v_{1}v_{2})$ scalings predicted in RHIC energy are also
investigated.

Anisotropic flows are defined as different $n$th harmonic
coefficients $v_n$ of the Fourier expansion for the particle
invariant azimuthal distribution,
\begin{equation}
\frac{dN}{d\phi}\propto{1+2\sum^\infty_{n=1}{v_n\cos(n\phi)}},
\end{equation}
where $\phi$ is the azimuthal angle between the transverse
momentum of the particle and the reaction plane. The anisotropic
flows $v_n$ can further be expressed in terms of single-particle
averages,
\begin{equation}
v_1 = \langle cos\phi \rangle = \langle \frac{p_x}{p_t} \rangle,
\end{equation}
\begin{equation}
 v_2 = \langle cos(2\phi) \rangle = \langle
\frac{p^2_x-p^2_y}{p^2_t} \rangle,
\end{equation}
\begin{equation}
 v_3 = \langle cos(3\phi) \rangle = \langle
\frac{p^3_x-3p_{x}p^2_y}{p^3_t} \rangle,
\end{equation}
\begin{equation}
 v_4 = \langle cos(4\phi) \rangle = \langle
\frac{p^4_x-6p^2_{x}p^2_y+p^4_y}{p^4_t} \rangle,
\end{equation}
where $p_x$ and $p_y$ are, respectively, the projections of
particle transverse momentum parallel and perpendicular to the
reaction plane, and $p_t$ is the transverse momentum ($p_t =
\sqrt{p_x^2+p_y^2}$).

\section{Theoretical framework}
The intermediate energy heavy-ion collision dynamics is complex
since both mean field and nucleon-nucleon collisions play the
competition role. Furthermore, the isospin-dependent role should
be also incorporated for asymmetric reaction systems.
Isospin-dependent quantum molecular dynamics model (IQMD) has been
affiliated with isospin degrees of freedom with mean field and
nucleon-nucleon collisions\cite{J. Aichelin,Y. G. Ma1,ZhangFS,J.Y.
Liu,Y. B. Wei,Y. G. Ma3}. The IQMD model can explicitly represent
the many-body state of the system and principally contains
correlation effects to all orders and all fluctuations, and can
well describe the time evolution of the colliding system. When the
spatial distance $\Delta r$ is smaller than 3.5\,fm and the
momentum difference $\Delta p$ between two nucleons is smaller
than 300\,MeV/$c$, two nucleons can coalescence into a
cluster\cite{J. Aichelin}. With this simple coalescence mechanism,
which has been extensively applied in transport theory, different
sized clusters can be recognized.

In the model the nuclear mean-field potential is parameterized as
\begin{equation}
U(\rho,\tau_{z}) = \alpha(\frac{\rho}{\rho_{0}}) +
\beta(\frac{\rho}{\rho_{0}})^{\gamma} +
\frac{1}{2}(1-\tau_{z})V_{c} \nonumber
\end{equation}
\begin{equation}
+ C_{sym} \frac{(\rho_{n} - \rho_{p})}{\rho_{0}}\tau_{z}+U^{Yuk},
\end{equation}
where $\rho_0$ is the normal nuclear matter density
($0.16$\,fm$^{-3}$), $\rho_n$, $\rho_p$ and $\rho$  are the
neutron, proton and total densities, respectively; $\tau_z$ is the
$z$th component of the isospin degree of freedom, which equals 1
or $-1$ for neutrons or protons, respectively. The coefficients
$\alpha$, $\beta$ and $\gamma$ are the parameters for nuclear
equation of state. $C_{sym}$ is the symmetry energy strength due
to the density difference of neutrons and protons in nuclear
medium, which is important for asymmetry nuclear matter ($C_{sym}
= 32$\,MeV is used). $V_c$ is the Coulomb potential and $U_{Yuk}$
is Yukawa (surface) potential. In the present work, we take
$\alpha $ = 124\,MeV, $\beta$ = 70.5\,MeV and $\gamma$ = 2 which
corresponds to the so-called hard EOS with an incompressibility of
$K$ = 380\,MeV.

\section{Results and discussions}
Now we move to the calculations. About $200,000$
$^{40}$Ca+$^{40}$Ca collisions have been simulated with hard EOS
at 25\,MeV/nucleon. In this study, we extract the physical results
at 200\,fm/$c$ for light charged particles when the system has
been in freeze-out.
 \begin{center}
\vspace{-0.5truein}
\includegraphics[width=8.5cm]{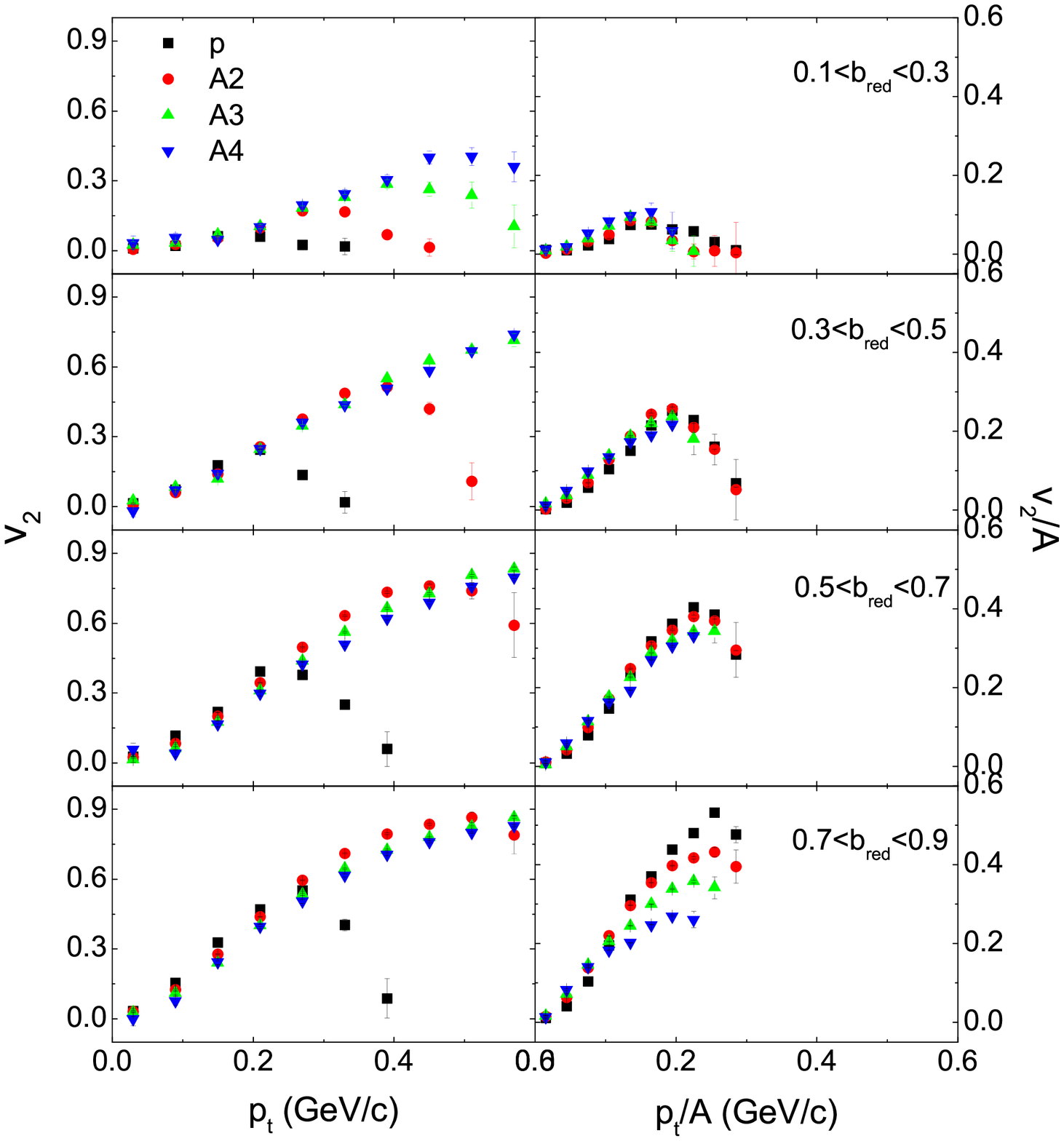}
\vspace{-1.2truein} \figcaption{\label{fig1} The transverse
momentum dependence of elliptic flow and the transverse momentum
per nucleon dependence of elliptic flow per nucleon for light
charged particles from 25 MeV/A $^{40}$Ca + $^{40}$Ca collisions
with different reduced impact parameters. Squares represent for
protons, circles for fragments of A=2, up-triangles for A=3,
down-triangles for A=4.}
\end{center}

Figure 1 shows the transverse momentum dependence of elliptic
flows (left column) and the transverse momentum per nucleon
dependence of elliptic flow per nucleon (right column) for light
charged particles from 25 MeV/A $^{40}$Ca + $^{40}$Ca collisions
with different reduced impact parameters. The reduced impact
parameter is defined as $b_{red} = b/b_{max}$ and $b_{max} = R_{p}
+ R_{t}$, where $R_{p}$ and $R_{t}$ are the radius of projectile
and target respectively. The four rows are for four different
reduced impact parameter bins of $0.1<b_{red}<0.3$,
$0.3<b_{red}<0.5$, $0.5<b_{red}<0.7$ and $0.7<b_{red}<0.9$,
respectively. Squares represent for protons, circles for fragments
of A=2, up-triangles for A=3, down-triangles for A=4. From the
figures of the transverse momentum dependence of elliptic flows
(left panel), they show that the elliptic flow is positive, and it
increases with the increasing $p_t$ and then becomes to decrease
with the increasing $p_t$ at a certain $p_t$, and the heavier
fragment has a greater $p_t$ of the inflection point. But in the
figures of the transverse momentum per nucleon dependence of
elliptic flow per nucleon (right panel), the curves for different
particles overlap with each other. This behavior is apparently
similar to the number of constituent quarks scaling of elliptic
flow versus transverse momentum per constituent quark ($p_t/n$)
for mesons and baryons which was observed at RHIC \cite{J.Adams}.
We called it the number of nucleons scaling of elliptic
flow\cite{Yan}, which reflects that the formation of the fragments
during the reaction obeys the coalescence mechanism. Figure 1 also
shows that the elliptic flow increases with impact parameter,
which indicates that the fragments prefer more to be emitted in
the reaction plane with greater eccentricity at larger impact
parameter. But the number of nucleons scaling is broken for
fragments with great $p_{t}/A$ at large impact parameter of
$0.7<b_{red}<0.9$, which may indicates that the collective effect
on lighter fragments are much stronger than heavier fragments,
i.e., lighter fragments are emitted at a higher thermal pressure
in the overlap zone.

\begin{center}
\vspace{-0.5truein}
\includegraphics[width=8.5cm]{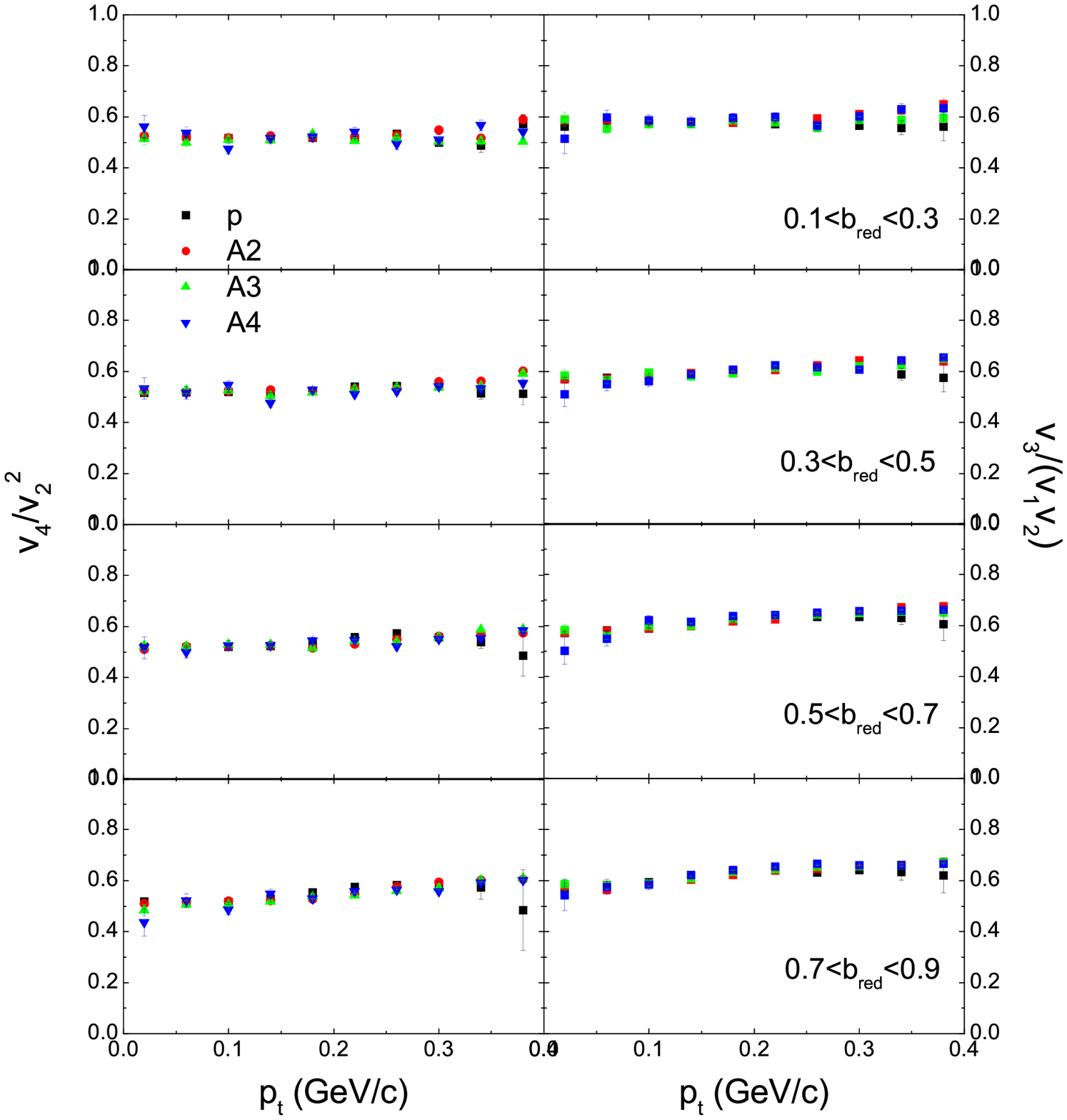}
\vspace{-1.2truein} \figcaption{\label{fig2} Transverse momentum
dependence of $v_4/v_2^2$ and $v_3/(v_1v_2)$ for light charged
particles from 25 MeV/A $^{40}$Ca + $^{40}$Ca collisions with
different reduced impact parameters. The symbols are same to that
of Fig.1.}
\end{center}

The RHIC experimental data also demonstrated a scaling
relationship between $v_4$ and ($v_{2}^{2}$) \cite{STAR03}. It has
been shown that such scaling relation follows from a naive quark
coalescence model \cite{Kolb2,ChenLW,Molnari} that only allows
quarks with equal momentum to form a hadron. Denoting the meson
anisotropic flows by $v_{n,M}(p_t)$ and baryon anisotropic flows
by $v_{n,B}(p_t)$ , Kolb\cite{Kolb2} et al. found that if quarks
have no higher-order anisotropic flows than the fourth term, one
can show that
$\frac{v_{4,M}}{v_{2,M}^{2}} \approx \frac{1}{4}+\frac{1}{2}
\frac{v_{4,q}}{v_{2,q}^{2}}$
and
$\frac{v_{4,B}}{ v_{2,B}^2} \approx \frac{1}{3} + \frac{1}{3}
\frac{v_{4,q}}{v_{2,q}^2}$,
where $v_{n,q}$ denotes the quark anisotropic flows. The meson and
baryon anisotropic flows thus satisfy the scaling relations if the
quark anisotropic flows also satisfy such relations, and this
ratio is experimentally determined to be 1.2 \cite{star_workshop}.
In view of the above behaviors of the flows at RHIC energies, we
display the anisotropic flows in intermediate energy. The left
panel of Figure 2 show the transverse momentum dependence of
$v_4/v_2^2$ for light particles from 25 MeV/A $^{40}$Ca +
$^{40}$Ca collisions with different reduced impact parameters. It
shows that the ratios of $v_4/v_2^2$ for different fragments are
nearly a constant value 0.5 at all the impact parameter bins. If
we assume the scaling laws of mesons and baryons are also valid
for A = 2 and 3 nuclear clusters, respectively, then $v_4/v_2^2$
for A = 2 and 3 clusters indeed give the same value of 1/2 as
nucleons(protons). Coincidentally the predicted value of the ratio
of $v_4/v_2^2$ for hadrons is also 1/2 if the matter produced in
ultra-relativistic heavy ion collisions reaches to thermal
equilibrium and its subsequent evolution follows the laws of ideal
fluid dynamics \cite{Bro}. It is interesting to note the same
ratio was predicted in two different models at very different
energies, which is of course worth to be further investigated in
near future. In addition, Kolb et al. suggested another scaling
relationship between $v_3$ and $v_1v_2$ as also insinuated by the
coalescence model.The right panel of Fig.2 display the $p_t$
dependence of the ratio $v_3/(v_1v_2)$. It shows that the ratios
of $v_3/(v_1v_2)$ for different LCP are also scaled and nearly a
constant value 0.6. It may be another parameter that can reflect
the thermalization of the matter in the overlap zone. It is also
worth to be further studied theoretically and experimentally at
both RHIC energy and intermediate energy.

\begin{center} \vspace{-0.5truein}
\includegraphics[width=8.5cm]{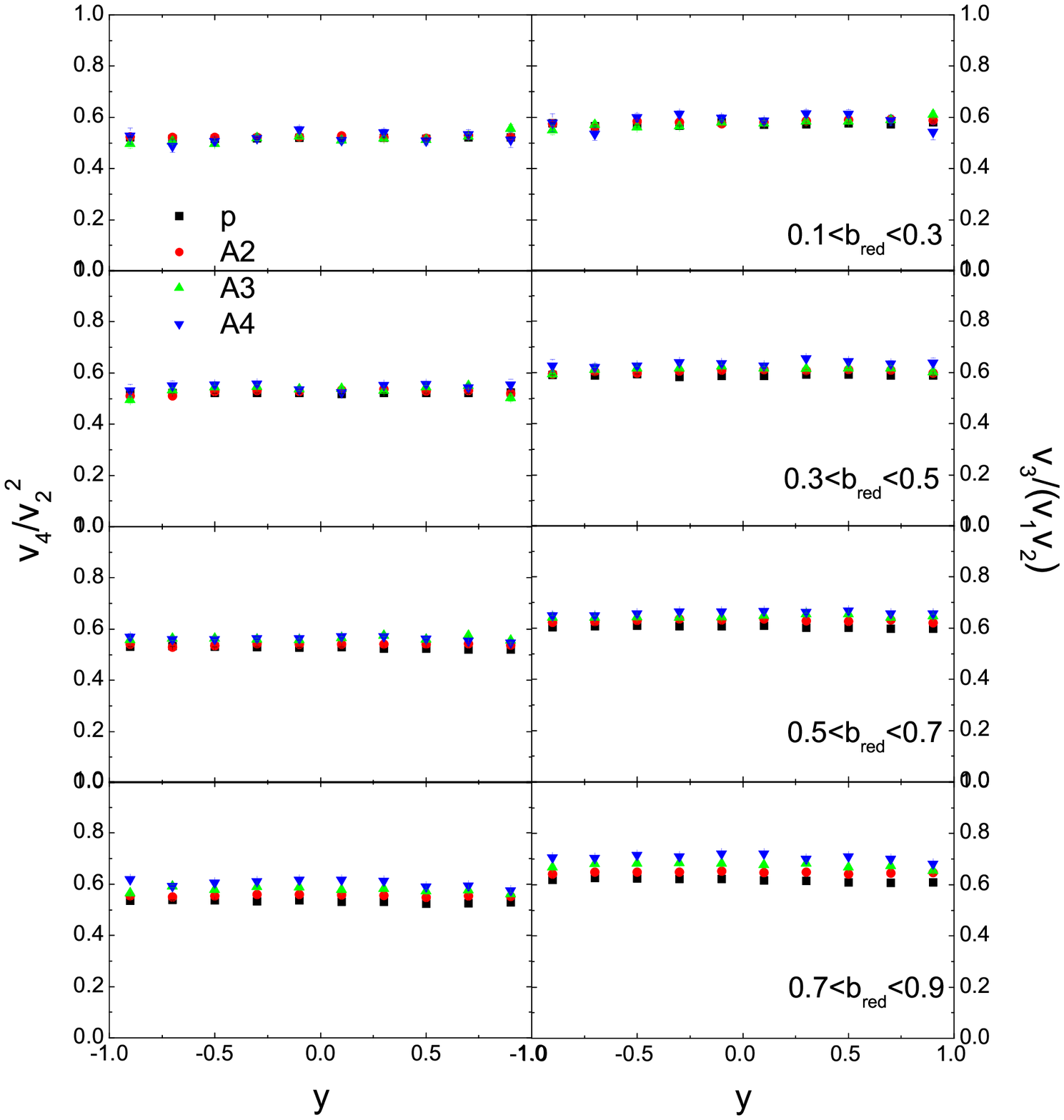}
\vspace{-1.2truein} \figcaption{\label{fig3} Rapidity dependence
of $v_4/v_2^2$ and $v_3/(v_1v_2)$ for light charged particles from
25 MeV/A $^{40}$Ca + $^{40}$Ca collisions with different reduced
impact parameters. The symbols are same to that of Fig.1.}
\end{center}

Figure 3 shows rapidity dependence of $v_4/v_2^2$ and
$v_3/(v_1v_2)$ for light charged particles from 25 MeV/A $^{40}$Ca
+ $^{40}$Ca collisions with different reduced impact parameters,
which are integrated for fragments with all possible $p_t$. So the
ratios of $v_4/v_2^2$ and $v_3/(v_1v_2)$ are as expected almost a
constant of 0.5 and 0.6 at different rapidity, respectively. But
to the peripheral collisions of $0.7<b_{red}<0.9$, the heavier
fragments have a little greater ratio value than the lighter ones.
That may because the flow difference of the light particles at
high $p_{t}/A$ as shown in the Figure 1. It may reflect that the
overlap zone is more approaching to the thermal equilibrium at the
moment the lighter fragments emitted for the peripheral
collisions.

\section{Summary}
We have investigated the scaling behaviors of elliptic flows and
the ratios of $v_4/v_2^2$ and $v_3/(v_1v_2)$ for light charged
particles at different collision parameters for the simulations of
$^{40}$Ca + $^{40}$Ca at 25 MeV/nucleon by the IQMD model. It is
shown that the number of nucleons scaling of elliptic flow is
existed at broad reduced impact parameter except for some
deviation at the peripheral collisions ($0.7<b_{red}<0.9$). The
ratios of $v_4/v_2^2$ and $v_3/(v_1v_2)$, which may reflect the
degree of thermalization of the matter produced in these heavy ion
collisions, are almost constant of about 0.5 and 0.6 respectively,
for all light fragments at different transverse momentum, rapidity
and impact parameter except for some deviation at the peripheral
collisions. It may reflect that at the peripheral collisions
different fragments may emitted at different thermalization extent
of the overlap zone because of the so large eccentricity. Further
theoretical and experimental investigations are awaiting.
\end{multicols}

\vspace{-1mm} \centerline{\rule{80mm}{0.1pt}} \vspace{2mm}

\begin{multicols}{2}
{}
\end{multicols}

\clearpage

\end{document}